\documentclass{ws-ijmpd}
\usepackage[super,compress]{cite}
\usepackage{multicol}

\usepackage{subfigure}
\usepackage[usenames]{xcolor}

\begin{document}

\markboth{M. Rashki, M. Fathi, B. Mostaghel , S. Jalalzadeh  }
{Interacting Dark Side of Universe Through Generalized Uncertainty Principle}

%
\catchline{}{}{}{}{}
%

\title{Interacting Dark Side of Universe Through Generalized Uncertainty Principle} 

\author{M. Rashki\footnote{mahdirashki@uoz.ac.ir, rashki.mahdi@gmail.com } }

\address{Department of Physics, Faculty of Science,  University of Zabol, Zabol, Iran. }
\author{M. Fathi}

\address{ Department of Physics, Shahid Beheshti University, G. C., Evin, Tehran, 19839, Iran.}

\author{B. Mostaghel}
\address{ Department of Physics, Shahid Beheshti University, G. C., Evin, Tehran, 19839, Iran.}

\author{S. Jalalzadeh\footnote{shahram.jalalzadeh@unila.edu.br}}
\address{ Federal University of Latin-American Integration,
 Technological Park of Itaipu PO box 2123, Foz do Igua\c{c}u-PR,  85867-670, Brazil, \\ Departamento de F\'isica, Universidade Federal de Pernambuco, Recife, PE 52171-900, Brazil.  }

\maketitle

\begin{history}
\received{Day Month Year}
\revised{Day Month Year}
\end{history}

\begin{abstract}
We investigate the impact of the generalized uncertainty principle proposed by some approaches to quantum gravity such as string theory and doubly special relativity on the cosmology. Using generalized Poisson brackets, we obtain the modified Friedmann and Raychaudhuri equations and suggest a dynamical dark energy to explain the late time acceleration of the Universe. After considering the interaction between dark matter and dark energy, originated from the minimal length, we obtain the effective cosmological parameters and equation of state parameter for dark matter and dark energy. Finally, we show that the resulting model is equivalent to the Phantom and Tachyon fields.
\end{abstract}


\ccode{PACS numbers: 04.50.Kd, 95.30.Sf, 95.35.+d, 98.80.-k }


\section{Introduction}	

The existence of minimal length  \cite{Mead1964ac}  has been predicted by various theoretical models such as string theory \cite{gross1987high} and Doubly Special Relativity  (DSR) \cite{Amelino}.  The presence of this minimal length may arise from imposing the quantum gravity effects in quantum mechanics through modification of the Heisenberg uncertainty principle, which is well-known as the generalized uncertainty principle (GUP)   \cite{Amati1988tn,Hossenfelder2003jz}. From the perturbative string theory, this length is due to the fact that the strings cannot influence through distances smaller than their size.  An interesting property of the existence of the minimal length is the modification of the standard commutation relation between canonical position and momentum in quantum mechanics and also Poisson brackets algebra in classical mechanics.

{Various authors have been used the deformed Poisson brackets (DPB) in classical cosmological models \cite{Benczik}. Also,  the quantum commutator of GUP is applied for a variety of quantum cosmological models \cite{QQ}. We should note that GUP deformation of Poisson brackets, for a point particles, implies a clear violation of the equivalence principle (EP) \cite{Farag1}. Consistency between DPB and EP can be partially recovered only for composite systems \cite{Tkachuk2012tk} and at the price of defining different $\beta$ parameters for different species of particles. That is expected that the EP to be violated to some degree by any quantum theory of gravity and by many alternative classical theories as well. The violation of this principle in scalar-tensor theories of classical gravity is linked to the existence of scalar gravitational fields in addition to the usual tensor field. On the other hand, one can expect the violation of EP in classical cosmological models at the presence of the deformed PBs. In fact,  the origin of this violation is in the usage of the minimal length of quantum gravity in the context of the classical gravitational theory. }
\par
Astrophysical observations show that the Universe is in an accelerating expansion phase \cite{Balbi2000tg} which is an evidence to the existence of dark energy. There exist several proposals which explain the origin of dark energy such as cosmological constant,  the expectation value of the vacuum ground state, which leads to the hierarchy problem \cite{Weinberg1988cp}. Modified theories of gravity such as $f(R,T)$ \cite{Shabani12014zza}, DGP braneworld \cite{Dvali2000hr}, the extrinsic cosmology \cite{Rostami2015} where  dark energy could actually be the manifestation of the local extrinsic shape of the spacetime, scalar-tensor theories \cite{Amendola1999qq,Boisseau},  modified matter models such as quintessence \cite{Fujii1982ms,Starobinsky},  K-essence, coupled dark energy \cite{Chiba2000im}  and dynamical dark energy models \cite{Alam2004jy} are some examples. The simplest candidates to explain the dynamical dark energy are the minimal coupling scalar fields like as the
quintessence and Phantom models \cite{Caldwell1997ii}.
\par
In this article,  using the classical GUP, we modify the Friedmann and Raychaudhuri equations to propose a new way to explain the interaction of the dark side of Universe. By using the reconstruction approach to scalar fields \cite{Alam2004jy}, we obtain a Phantom potential and Tachyon field in the model and show that this potential originated from the GUP effects. In Section II, we present the model that assists in our investigation. In section III we consider an interaction between dark energy  and dark matter trough GUP, and finally in the last two sections we reconstruct Phantom and Tachyon fields which are consistent to the model.

\section{Classical cosmology at the presence of GUP}
Let us consider  the   isotropic and homogeneous Friedmann-Lema\^itre-Robertson-Walker (FLRW) Universe with line element
 \begin{eqnarray}\label{line}
  ds^2= -N^2 (t)d{t} ^2 +{a^2 (t)} \left(\frac{dr^{2}}{{1-k r^{2}}}+r^{2} d\Omega^2\right),
  \end{eqnarray}
  where $N(t)$ denotes the lapse function, $a(t)$ is the scale factor of the Universe  and $k=0,+1,-1$  represents the usual
spatial curvature. The action functional corresponding to the line element
(\ref{line}) displays in the gravitational and matter sectors
(with the latter as perfect fluid) \cite{haw-ellis}
\begin{align} \label{functional}
\begin{split}
S&=\frac{1}{16\pi G} \int{\sqrt{-g}{R} d^4x}-\int {\sqrt{-g} \rho  d^4 x}
\\
&=\frac{3\pi \sigma}{4G} \int(-\frac{a\dot{a}^2}{N}+{k} Na)dt-2\pi^2 \sigma \int{Na^3}\rho dt,
\end{split}
\end{align}

where $\rho=\sum\limits_{i}\rho_{i}$ is the total matter density  of the Universe,  $2\pi^2 \sigma$ is the spatial volume of the metric which remains finite for the observable Universe with the radius about $ H_{0}^{-1}$  $(\sigma=1\,\, \text{for}\,\,k=1)$ and overdot  denotes differentiation  respect to cosmic time $t$.
According to  the energy conservation  the  matter density  content of the   Universe  takes the form $\rho_{i}=\rho_{0i}(a/a_{0})^{-3(w_{i}+1)}$  where $w_{i}$ denotes the equation of state (EoS), $w_{i}={p_{i}}/{\rho_{i}}$,  for $i$-th component of fluid and $\rho_{i0}$ is the present density.
The action functional of the model will be
 
\begin{equation}
S=\frac{3\pi \sigma  }{4G}\int{\left[\left(\frac{-\dot{a}^{2}a}{N}+k Na\right)-N a^{3}\sum\limits_{i}\rho_{0i}(\frac{a}{a_{0}})^{-3(\omega_{i}+1)}\right]}dt\,.
\end{equation}

One can rewrite the present density of $i$-th component of density, $\rho_{0i}$, in terms of density  parameter, $\Omega_i$, as $\rho_{0i}=\dfrac{3{H}^{2}_{0}\Omega_{i}}{8\pi G}$ . If we use the dimensionless time defined by $d\eta=H_0dt$
and ignoring the overall constant factor, Lagrangian of the model in terms of density parameters will be
\begin{eqnarray}\label{lag. density}
\mathcal{L}=-\frac{M}{2}\left(\frac{x\dot{x}^{2}}{N}+N \Omega_{k} x + N \sum\limits_{i}\Omega_{i} x^{-3w_{i}}\right)\,,
\end{eqnarray}
where  $\Omega_k=\frac{-{k}}{a_0^2H_0^2}$,  $x=\frac{a}{a_{0}}$ denotes the rescaled dimensionless scale factor
\begin{eqnarray}\label{mass}
M=\frac{3\pi a_0^3H_0}{2G},
\end{eqnarray}
and after this overdot denotes differentiation respect to $\eta$.
The conjugate momentum  to the scale factor $x$ and the primary constraint  are given by
\begin{equation}\label{primary constraint}
p=\frac{\partial\mathcal{L}}{\partial\dot x}=-\frac{M x}{ N}\dot x,\hspace{.4cm}
p_{ N}= \frac{\partial\mathcal{L}}{\partial\dot {N}} =0.
\end{equation}

Consequently,  the Hamiltonian  corresponding to the Lagrangian (\ref{lag. density})
will be
\begin{equation}\label{canonical.hamilton}
{\mathcal H}= N\left[-\frac{p^2}{2Mx} +
\frac{M}{2}\Omega_{k} x +\frac{M}{2}\sum\limits_{i}\Omega_{i}x^{-3w_{i}}\right].
\end{equation}

In Eq.~\eqref{canonical.hamilton}, $N$ is a Lagrange multiplier, hence  it  compel the Hamiltonian constraint
\begin{eqnarray}\label{const.hamilton}
-\frac{p^2}{2M x} +\frac{M}{2}\Omega_{k} x +\frac{M}{2} {\sum\limits_{i}}\Omega_{i}x^{-3w_{i}} =0.
\end{eqnarray}
\subsection{The Generalized Uncertainty Principle}
  The generalized uncertainty principle between canonical conjugates $(X,\Pi_X)$ which leads to a minimum  length  in one dimension is given by \cite{Pedram2010zz}
 \begin{eqnarray}\label{U P }
\triangle X\triangle \Pi_{X} \geqslant\frac{1}{2}\left(1+\beta(\triangle \Pi_{X})^{2}+\gamma
\right),
\end{eqnarray}
where $\beta$ and $\gamma$ are two parameters,  independent of $ \Delta{X}$ and $\Delta{\Pi_{X}}$, but may, in general, depend on the expectation value of the position and the momentum operators. The above generalization of uncertainty principle corresponds to the following commutation relation
\begin{eqnarray}\label{commutation relation}
[X,\Pi_{X}]= i(1+\beta \Pi_{X}^{2}),
\end{eqnarray}
where by comparing Eqs.~\eqref{U P } and \eqref{commutation relation}, we obtain $\gamma=\beta\langle\Pi_X\rangle^{2}  $.
  A more general case of such commutation relations are studied in Ref.~[\refcite{Hinvichsen1996}]. The various applications of the low energy effects of the modified Heisenberg uncertainty principle relations have been extensively studied in Refs.~[\refcite{Brau,Akhoury2003,Hofmann2003,Nozari2006}]. The above equations show that the smallest uncertainty in position has the value
\begin{eqnarray}\label{man1}
\Delta X_{min}=\sqrt\beta\,,
\end{eqnarray}
 wherein generally, the parameter of $\beta$  can't be fixed by the theory.
 \par
One major feature of Eq.~(\ref{man1}) is that the physics below $\sqrt\beta$ becomes inaccessible and therefore it defines a natural cut-off which prevents from the usual UV divergences. On the other hand, the second consequence of GUP is the appearance of an intriguing UV/IR mixing, first noticed in the ADS/CFT correspondence \cite{Sus}. As we know, the UV/IR mixing means that one can probe short distances physics (high energy physics) by long distances physics (low energy physics), and therefore, it justifies the use of classical mechanics in the presence of a minimal length. We point that the UV/IR mixing is also a feature of noncommutative quantum field theory \cite{Do}. On the other hand, some scenarios have been proposed where non zero minimal length is related to large extra dimensions \cite{HO1}, to the running coupling constant \cite{Ho2} and to the physics of black holes production \cite{Ho3}.
\par
Let us now investigate the effects of deformed Poisson algebra in the presence of a minimal length. Since there is a UV/IR mixing embodied in the deformed commutation relation \cite{Benczik}, it is also important to study the effects of the minimal length in a classical context. Note that to obtain the classical and commutative limit, it is necessary to consider simultaneously $\hbar\rightarrow0$ and $\beta\rightarrow0$. Specifically, if $\beta\neq0$ there is no classical limit. Therefore, to embedding the effect of UV-IR mixing of deformation in the classical cosmology at the presence of minimal length, it is enough to consider the limit of $\hbar\rightarrow0$. Then, the modified Poisson brackets corresponding to the GUP will be \cite{Nam}
\begin{align}\label{a11a}
&\{X,X\}=\{\Pi_X,\Pi_X\}=0,\\
&\{X,\Pi_X\}=1+\beta\Pi_X^2.
\end{align}

Such deformed Poisson algebra is used in Ref.~[\refcite{Benczik}] to investigate the effects of the deformation on the classical orbits of particles in a central force field and on the Kepler third law. Also, the cosmological constant problem and removability of initial curvature singularity are investigated in Ref.~[\refcite{Bina}]. The effect of modified Poisson algebra defined in Eq.~\eqref{a11a} is also investigated in a dilatonic cosmological model at Ref.~[\refcite{Vakili1}] and in Bianchi I and II cosmological models in Ref.~[\refcite{Farag}]. It is examined in de Sitter and Anti-de Sitter classical cosmological models in Ref.~[\refcite{Vakili2}]. The effect of above algebra on inflation parameters is investigated in Ref.~[\refcite{Nozari}]. In addition, the authors of  Ref.~[\refcite{Zeynali}] are studied the effects of the deformed Poisson algebra Eq.~\eqref{a11a} in multidimensional cosmology. And as the last example, we should mention, the signature changing classical cosmology at the presence of minimal length is described in Ref. [\refcite{Ghaneh}].
The result of the above investigations is that the modifications of physics at small scales, via introducing the minimal length in deformed Poisson algebra, have a rather profound effect on classical physics at
large scales, something similar to the UV/IR mixing.

One can easily show that the deformed Poisson bracket algebra given in Eq.~\eqref{a11a} is satisfied by the classical dynamical position $ x $ and conjugate momentum $p $  given by  \cite{Bernardo2015}
\begin{equation}\label{transform}
X=x,\hspace{.3cm}
\Pi_{X}=p+\frac{\beta}{3}p^{3},
\end{equation}

in which $x$  and  $p$
 obey the usual Poisson bracket $\{x, p\}=1$. Note that in the natural units
 that we used in this article all quantities are dimensionless: $[S]=[\mathcal
 L]=[x]=[p]=[M]=[\beta]=1$.
 
By substituting  relations (\ref{transform}) into  Eq.(\ref{canonical.hamilton}) the  deformed Hamiltonian will be
\begin{align}\label{deformed const.hamilton}
{\mathcal H}_{def}=N\left (-\frac{1}{2Mx}\left[ p^{2}+\frac{2\beta}{3}p^{4}
\right] +\frac{M}{2}\Omega_{k}x +\frac{M}{2} \sum\limits_{i}\Omega_{i}x^{-3 w_{i}}\right ).
\end{align}

Consequently, according to Eq.(\ref{deformed const.hamilton}) the equations of motion  in the  gauge of $N=1$ will be
\begin{align}\label{equation of motion}
&\dot{x}= \dfrac{\partial{{{\mathcal{H}}_{def}}}}{\partial{p}}=-\dfrac{p}{Mx}{\left(1+\frac{4\beta p^{2}}{3}\right)}\,\\
&\dot{p} =-\dfrac{\partial{{\mathcal{H}}_{def}}}{\partial{x}}=-
\dfrac{p^2}{2 Mx^{2}}\left(1 +\dfrac{2\beta}{3}p^{2} \right)+\dfrac{M}{2}\Omega_{k}
-\dfrac{3M}{2}  \sum \limits_{i} \Omega_{i}w_{i}x^{-3w_{i}-1}.
\end{align}

  The first equation in (\ref{equation of motion}) gives us the relation
of momentum with $\dot x$. By inserting the obtained momentum in Hamiltonian
constraint (\ref{deformed const.hamilton}) the  deformed Friedmann equation will be
\begin{align}\label{friedmann}
\left(\frac{H}{H_0}\right)^2=\Omega_{k}x^{-2}+
  \sum\limits_{i}\Omega_i x^{-3(w_{i}+1)} +2\beta M^2 x^4\left(\Omega_{k}x^{-2}+
   \sum\limits_{i}\Omega_{i} x^{-3(w_{i}+1)} \right)^{2},
\end{align}

where $H=\frac{1}{a}\frac{da}{dt}$ denotes the Hubble parameter in the comoving
coordinates. The second equation of (\ref{equation of motion}) gives the modified  Raychaudhuri equation
\begin{align}\label{raychaudhuri}
\begin{split}
\frac{1}{H_0^2}\frac{\ddot{x}}{x}=&-\frac{1}{2} \sum\limits_{i} \Omega_{i} (1+3w_{i})x^{-3(w_{i}+1)}
+2\beta M^2x^4\left(\frac{\Omega_k}{x^2}+\sum\limits_i\Omega_ix^{-3(w_i+1)}\right)\times
\\ &\left(\frac{\Omega_k}{x^2}-\sum\limits_i\Omega_iw_ix^{-3(w_i+1)}\right)\,.
\end{split}
\end{align}

Also, the age of Universe will be
\begin{align}\label{age1}
t_0=\frac{1}{H_0}\int_0^1{dx[\Omega_k+
  \sum\limits_i \Omega_i x^{-3w_{i}-1}+2 \beta
 M^2x^2 (\Omega_{k}+  \sum\limits_{i} \Omega_{i} x^{-3w_{i}-1})^{2}]^{-\frac{1}{2}}}.
\end{align}

Let us consider a spatially flat Universe dominated by vacuum energy with EoS $w=-1$. According to  Eq.(\ref{friedmann}) the modified Friedmann equation becomes
\begin{eqnarray}\label{mod. hamilton}
\begin{array}{cc}
{H}^{2}={H}_{0}^{2}\left(\Omega_{\Lambda}+2\beta M^2\Omega^2_{\Lambda}x^{4}\right)\equiv{H}_{0}^{2} \Omega_{{DE}}{(x)},
\end{array}
\end{eqnarray}
where we defined the  modified dark energy parameter   as
\begin{eqnarray}\label{dark density}
\Omega_{{DE}}(x)=\Omega_{\Lambda}+2\beta M^2\Omega^2_{\Lambda}x^4,
\end{eqnarray}
and  $\Omega_{\Lambda}$ denotes the density parameter of cosmological constant at the present epoch.
From Eq.(\ref{dark density}) we define
\begin{eqnarray}\label{omega beta}
\Omega_{\beta}^{0}\equiv 2\beta M^2\Omega^2_{\Lambda}=\Omega^{0}_{{DE}}-\Omega_{\Lambda}\,,
\end{eqnarray}
which states the dependence of GUP parameter $\beta$ to the density parameters of dark energy and the cosmological constant. As we see in our model the GUP parameter could take the positive and negative value according to the difference between $\Omega_{DE}^{0}$ and $\Omega_{\Lambda}$.{ With substituting value of $M$  from Eq. \eqref{mass} in Eq.\eqref{omega beta}  we obtain the order of magnitude of $\beta$ as
\begin{eqnarray}
\beta=\frac{\Omega_{\beta}}{2M^{2}{\Omega_{\Lambda}}^{2}}\approx \frac{G^{2}}{a_{0}^{6} H_{0}^{2}}\approx \frac{G^{2}}{d_{(H_0)}^{6} H_{0}^{2}} \approx 10^{-84} \,,
\end{eqnarray}
where $d_{H_{0}} $ denotes the Hubble distance at the present epoch and $\Omega_{\beta}$ is in order of $10^{-2}$. The order of the GUP parameter is reasonable in our model because we apply the GUP approach to the evolution of the Universe in the present epoch where the quantum effects are small. Note that this limit is 49 orders of magnitude below the Planck scale. } 
\par
{Benczik and et.al in Ref.~[\refcite{Benczik}] obtain the bound $\beta\lesssim 10^{-66}$ from investigation the precession perihelion of Mercury by GUP-deformed Newtonian mechanics. While Scardigli and et.al in Ref.~[\refcite{Scardigli2}] compute upper bound for the GUP parameter in light deflection by deformation Schwarzschild metric as $\beta \lesssim 10^{78}$, and for perihelion precession by that deformed metric obtained the lower bound for GUP parameter as $\beta \lesssim 10^{69}$. As well as, in Ref.~[\refcite{Farag Ali}] the author investigated derivation of the modified Hawking temperature from the GUP, and find an upper bound on GUP parameter for the deflection of light $\alpha_{0}<10^{41}$, which there is $\alpha_{0}$ in contrast to Ref.~[\refcite{Scardigli2}]  namely $ \alpha_{0}^{2}\sim \beta $,  and upper bound for of the perihelion precession of Mercury is $\alpha_{0}<10^{35}$.
Recently Scardigli et.al in Ref.~[\refcite{Scardigli3}] compute the value of GUP parameter by comparing two different low energy corrections. They, first involved $\beta$ through the GUP procedure and then insert the corrections of Newtonian potential which is leads to a quantum correction for Schwarzschild metric that it computed by Donoghue\cite{Donoghue}. From this approach, they obtain that the order of GUP parameter is $\beta=\frac{82\pi}{5}$. } 
\par

  The dynamic of dark energy in our model originates from the parameter $\beta $.

 We can write the dark energy density parameter in terms of red-shift as
\begin{eqnarray}\label{dep}
\Omega_{{DE}}(z)=\Omega_{\Lambda}+\Omega_{\beta}^{0}(1+z)^{-4}\,,
\end{eqnarray}
\par
\begin{figure}[t!]
 \centering
\subfigure[]{ \includegraphics[width=0.48\textwidth]{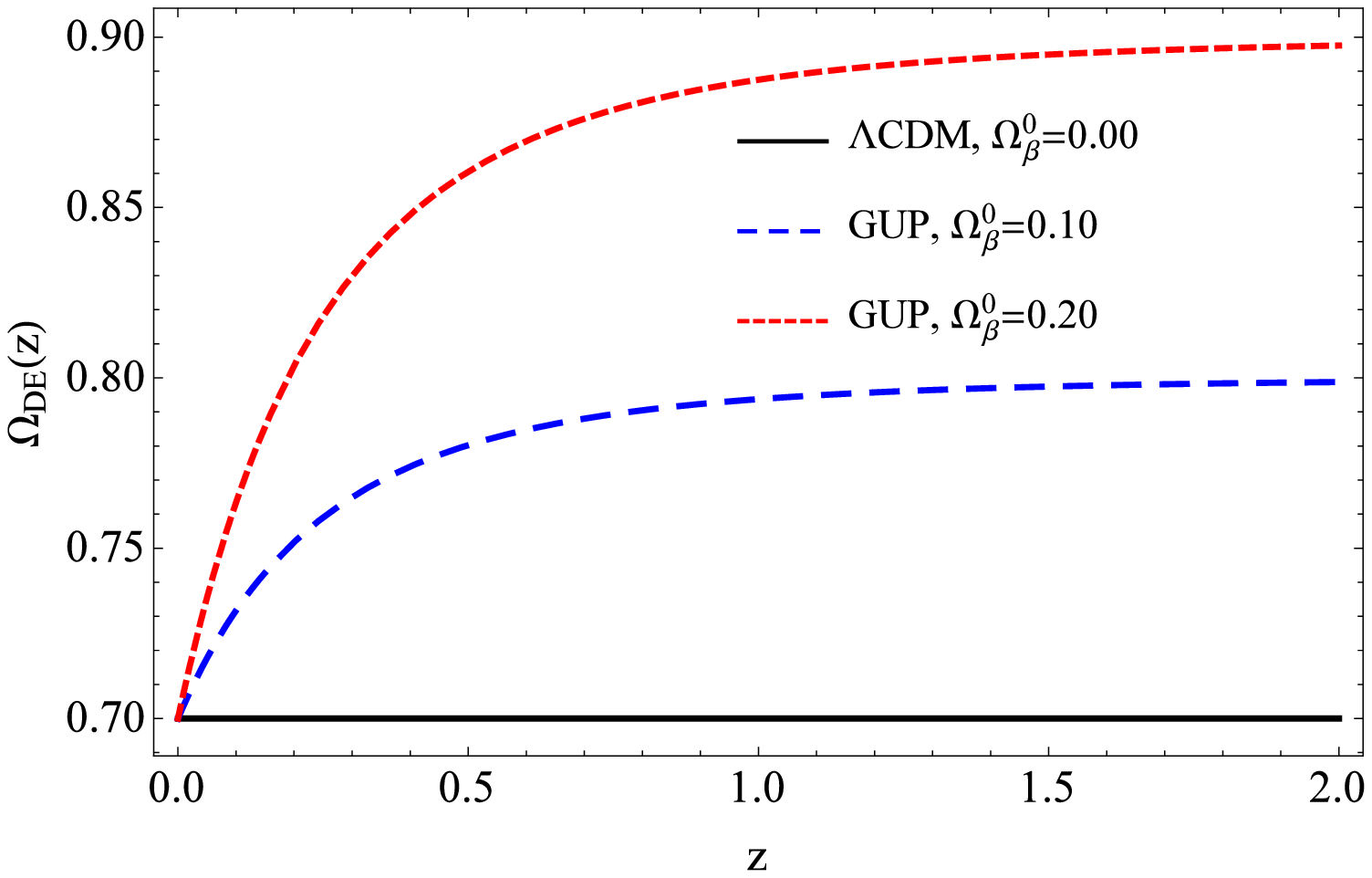}\label{fig:deplt}}
\subfigure[]{\includegraphics[width=0.48\textwidth]{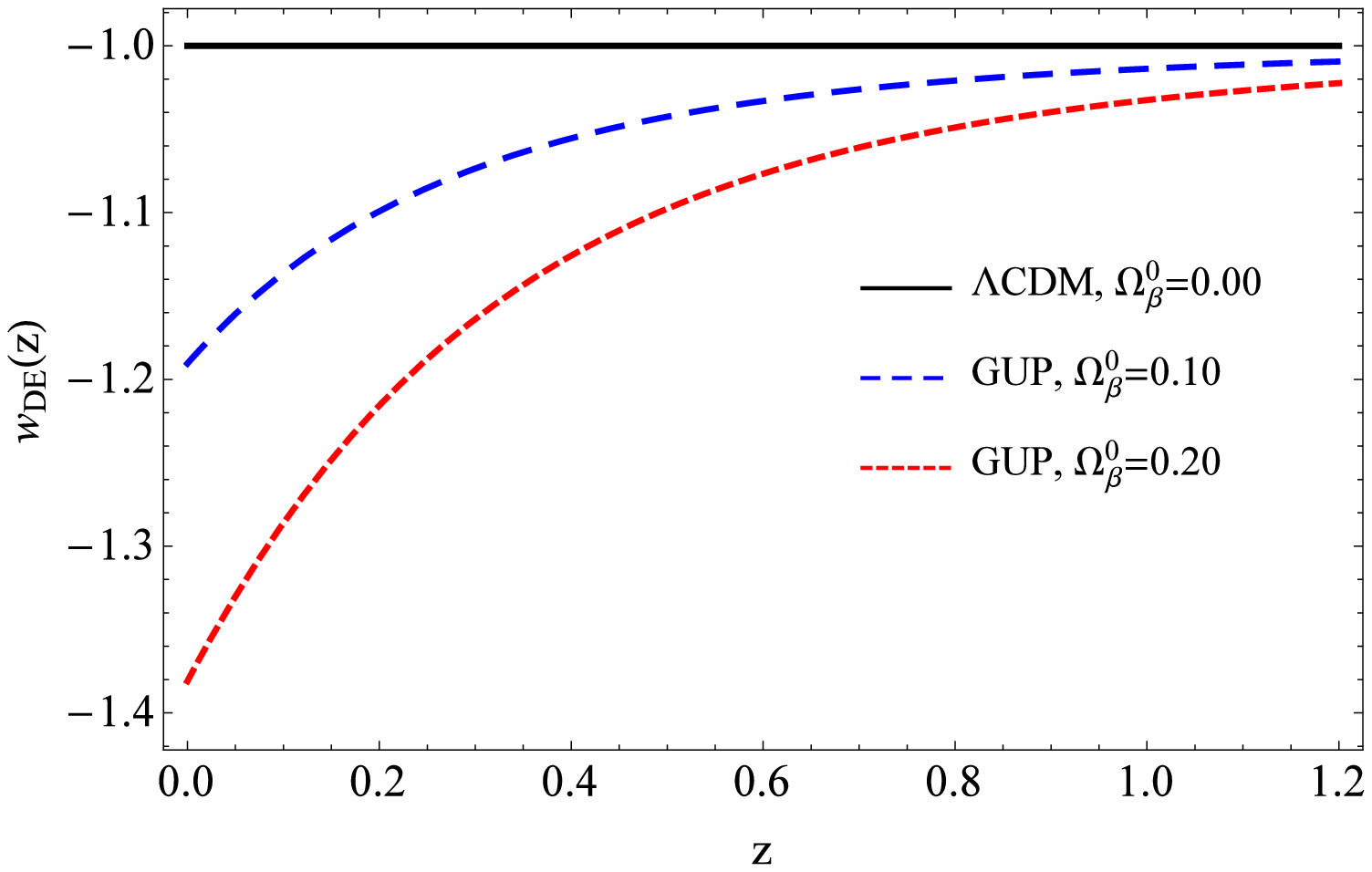}  \label{eos}}
 \caption{
\textit{panel a}: $\Omega_{{DE}}(z)$ as a function of redshift. \textit{panel b}:The evolution of EoS of dark energy  respect to red-shift. We change the value of $\beta$ while other parameters have been fixed to $\Omega^0_{DE}=0.70$ and $\Omega^0_{DM}=0.30$.
 }
\end{figure}

We plot the evolution of $\Omega_{{DE}}(z)$ in the Fig.~\ref{fig:deplt}. To make more sense, we compare the behavior of GUP model with standard $\Lambda$CDM model.  According to this figure, there is a meaningful difference between standard $\Lambda$CDM model and GUP dark energy.
The modified EoS for dark energy becomes
\begin{align}\label{D EOS}
w_{{DE}}=-1+\frac{1}{3} \frac{d \ln\Omega_{{{DE}}}(z)}{d \ln(1+z)}=-1-\frac{4}{3}\frac{\Omega^{0}_{\beta}(1+z)^{-4}}{\Omega^{0}_{\Lambda}+\Omega^{0}_{\beta}(1+z)^{-4}}.
\end{align}
This relation shows that in the early Universe, i.e., $z=\infty$,  we have $\Omega_{{DE}}\approx\Omega^0_{\Lambda}$ and the effect of the second term is important in the very late times. { In Fig.~\ref{eos}, the equation of state of the proposed model is plotted for two values of the $\beta$ parameter. }

According to Eq.~(\ref{D EOS}), at the present epoch, the modified EoS parameter of the dark energy is
\begin{eqnarray}\label{eos2}
w^{0}_{{DE}}=-1-\frac{4}{3} \frac{ \Omega^{0}_{{\beta}}   }{\Omega^{0}_{{DE}}}\,.
\end{eqnarray}
For the positive value of the parameter $\beta$, we get ${{\Omega^0_{\beta}\geq 0}}$ and consequently, ${w^0_{DE}<-1}$. It implies that in our approach, Universe enters  the Phantom phase at the late time.

\section{Interacting Dark Energy and Dark Matter}
Let us now consider the interaction between dark energy and the dark matter content of the Universe (where we neglect the contribution of baryonic matter {and radiation}) \cite{Berger2006db}. As we know the continuity equation for dark energy and dark matter in standard cosmology are
\begin{align}\label{continuity}
&\dot{\rho}_{\Lambda}=0,\\
&\dot{\rho}_{DM}+3H\rho_{DM}=0.
\end{align}
Substituting the effective dark energy parameter from Eq.~(\ref{dark density}) into  Eq.~(\ref{continuity}) we obtain the modified continuity equations which can be interpreted as the interaction between dark energy and dark matter. The simplest form of the interaction equations are
\begin{align} \label{interone}
&\dot{\rho}_{DE}=-3H(1+w_{DE})\rho_{DE}=Q,\\
&\dot{\rho}_{DM}+3H\rho_{DM}=-3Hw_{DM}=-Q,
\end{align}
where $w_{DE}$ and $w_{DM}$ denote the effective EoS parameters for dark energy and dark matter respectively and $Q$ represents the interacting term. Generally  {speaking, the function} $Q$ represents the energy transition between dark energy and dark matter.  Using  Eq.~(\ref{interone}), we obtain the following two equations  for EoS and energy density parameter of  dark matter content
\begin{align} \label{inter2}
&w_{DM}=-(1+w_{DE})\frac{\Omega_{DE}}{\Omega_{DM}},\\
&\frac{d\Omega_{DM}}{dz}-\frac{3}{1+z}\Omega_{DM}=-\frac{d\Omega_{DE}}{dz}  \label{inter2}.
\end{align}
 Substituting $\Omega_{DE}$ from Eq.(\ref{dep}) into the second Equation
of (\ref{inter2}) gives
 \begin{equation}\label{omegam}
\Omega_{DM}=\Omega_{DM}^0(1+z)^3+\frac{4}{7} \Omega_{\beta}^0\left((1+z)^3-(1+z)^{-4}\right),
\end{equation}

\begin{figure}[t!]
\centering
\subfigure[]{ \includegraphics[width=0.48\textwidth]{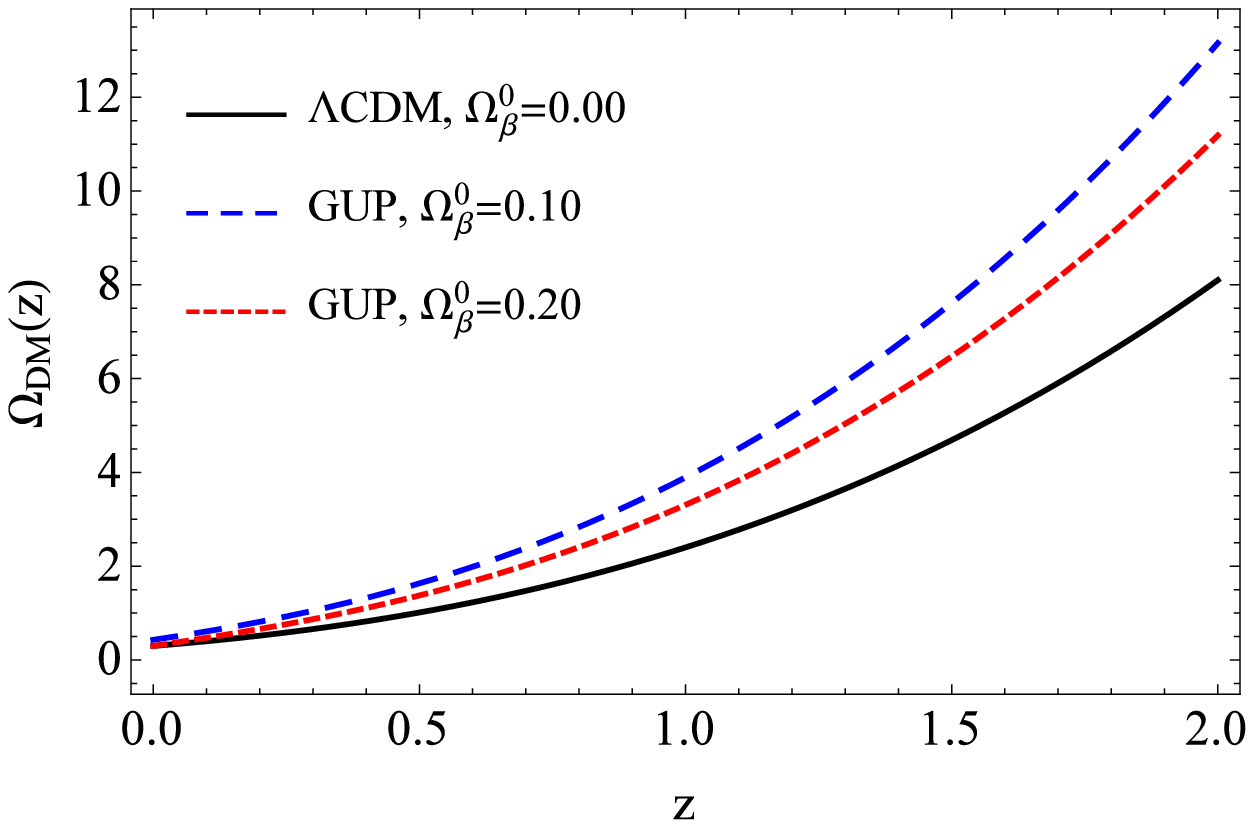}\label{11}}
\subfigure[]{ \includegraphics[width=0.48\textwidth]{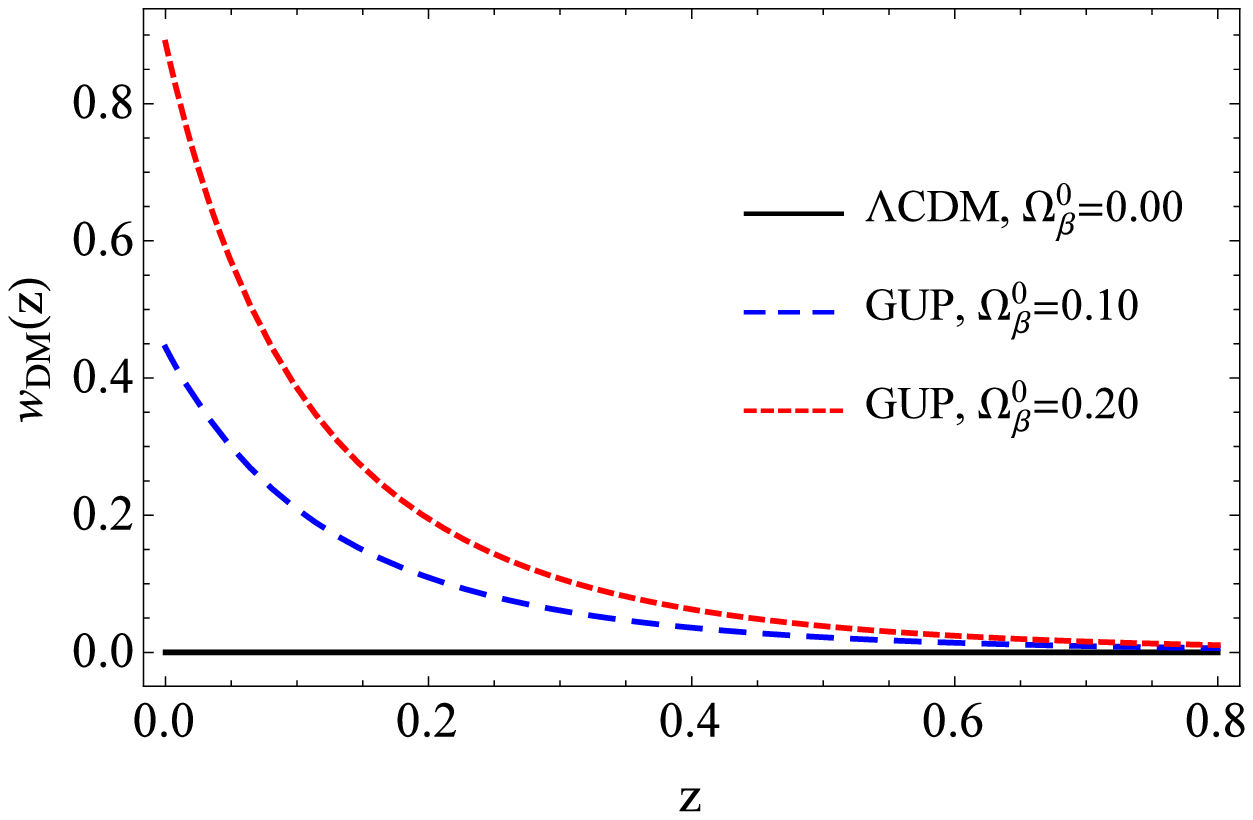}\label{dark matter eos}}
 \caption{\textit{panel a}: The evolution of the dark matter density parameter with respect to red-shift. \textit{panel b}: Equation of state of dark matter respect to the red-shift. We change the value of $\Omega^0_{\beta}$  while other parameters have been fixed to $\Omega^0_{DE}=0.70$ and $\Omega^0_{DM}=0.30$. }
\end{figure}

The evolution of interacting dark matter respect to the red-shift, shown in the Fig. (\ref{11}).
The effective EoS parameter for dark matter takes the form
\begin{align}\label{eosm}
\begin{split}
w_{DM}&=-1+\frac{1}{3}\frac{d \ln(\Omega_{DM})}{d \ln(1+z)}\\
&=\frac{28\Omega_{\beta}^{0}(1+z)^{-4}}{21\Omega_{DM}^0(1+z)^3+12\Omega_\beta^0((1+z)^3-(1+z)^{-4})}.
\end{split}
\end{align}

\begin{figure}
 \centering
 \includegraphics[width=0.5\textwidth]{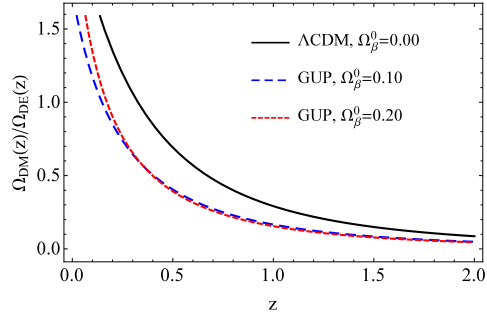}
  \caption{The ratio of dark energy density parameter over dark matter density parameter as a function of red-shift. This figure shows that for $\Omega^0_{\beta}>0$ the contribution of dark matter in the model is lower than the standard $\Lambda${CDM} model, while for $\Omega^0_{\beta}<0$ the contribution of dark matter in the model is more than a standard $\Lambda${CDM} model. We change the value of $\Omega^0_{\beta}$  while other parameters have been fixed to $\Omega^0_{DE}=0.70$ and $\Omega^0_{DM}=0.30$. }
 \label{fig:energyfraction}
\end{figure}

The evolution of   $w_{DM}$  respect to red-shift is shown in Fig.(\ref{dark matter eos}).
{For the positive value of the parameter $\Omega^0_{\beta}$, dark matter behaves like as a stiff fluid with $w^0_{DM}\approx  1$.}
When the GUP parameter,{ $\beta$ (and consequently $\Omega^0_{\beta}$),} tends to zero, the EoS parameter $w_{DM}$ and density parameter of dark matter, $\Omega_{DM}$, take the values zero and $\Omega_{DM}^0$   respectively which shows that the deviation of EoS for ordinary dark matter is a consequence of the GUP {and the special form of interaction in the Eq.~\eqref{inter2} for the proposed dark energy model.}
{To make more sense and in order to compare the role of the parameter $\Omega^0_{\beta}$ on the dynamics of Universe respect to the standard model, in the Fig.~\ref{fig:energyfraction} we plot the ratio of dark energy density parameter to the dark matter  density parameter, $\Omega_{{DM}}(z)/\Omega_{{DE}}(z)$. For the positive value of the parameter $\Omega^0_{\beta}$, the contribution of the dark matter in the Universe is lower than the standard $\Lambda$CDM model.}

\begin{figure}[t!]
\centering
\includegraphics[width=0.5\textwidth]{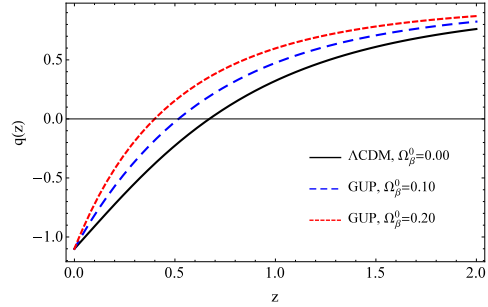}
\caption{The behavior of deceleration parameter with respect to the red-shift.
  We change the value of $\Omega^0_{\beta}$ while other parameters have been fixed to $\Omega^0_{DE}=0.70$ and $\Omega^0_{DM}=0.30$.}
\label{decel}
\end{figure}

Deceleration parameter defined by
\begin{eqnarray}
q(z)=\frac{(1+3w_{DM})\rho_{DM}+(1+3w_{DE})\rho_{DE}   }{2(\rho_{DM}+\rho_{DE})}.
\end{eqnarray}
{In Fig.~\ref{decel} we plot the deceleration parameter $q(z)$ of the proposed model respect to the red-shift. According to this figure, Universe enters to the late time acceleration expansion with a time delay, respect to the concordance $\Lambda$CDM model.
}
As we see in Fig.~\ref{decel}, for high red-shifts, the deceleration parameter tends to $1$ 
and the dynamic of Universe determined by dark matter component in our model.

\section{Reconstruction of Scalar Field}
\subsection{Reconstruction of Phantom Fields}
Now, we try to reconstruct a scalar field which coincides with results of the previous section. We use the reconstruction method which proposed in Refs.~\refcite{Tsujikawa2005ju,Guo2005ata}. 
{Since for different values of the parameter $\Omega^0_{\beta}$, there is no phantom crossing, we can consider scalar fields with minimal coupling \cite{Mostaghel:2016lcd}.}
  In minimal coupling with Einstein gravity, energy density and pressure for scalar fields are
\begin{eqnarray}
\begin{array}{cc}
\rho_{\phi}=\frac{1}{2}\epsilon\dot{\phi}^{2}+V(\phi),\\
P_{\phi}=\frac{1}{2}\epsilon\dot{\phi}^{2}-V(\phi),
\end{array}
\end{eqnarray}
{where $\epsilon=\pm1$ for quintessence and phantom field respectively} and hereafter we will set $\epsilon =-1$. Using the EoS parameter,  $w_{\phi}={P_{\phi}}/ {\rho_{\phi}}$, the kinetic term becomes
\begin{equation}
\dot{\phi}^{2}=\epsilon \left(1+w_{\phi}\right)\rho_{\phi}.
\end{equation}

\begin{figure}[t!]
\centering
\includegraphics[width=0.5\textwidth]{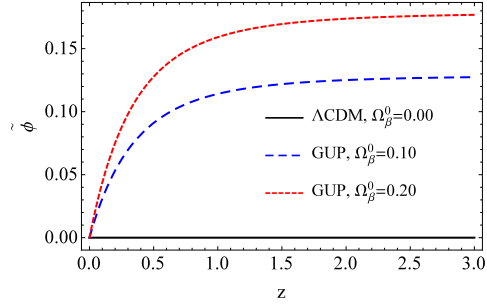}
\caption{Evolution of scalar field with respect to the red-shift.  We change the value of $\Omega^0_{\beta}$ while other parameters have been fixed to $\Omega^0_{DE}=0.70$ and $\Omega^0_{DM}=0.30$ and we set $\tilde{\phi}(z=0)=0$.}
\label{p-f}
\end{figure}

\begin{figure}[t!]
\centering
\subfigure[]{\includegraphics[width=0.48\textwidth]{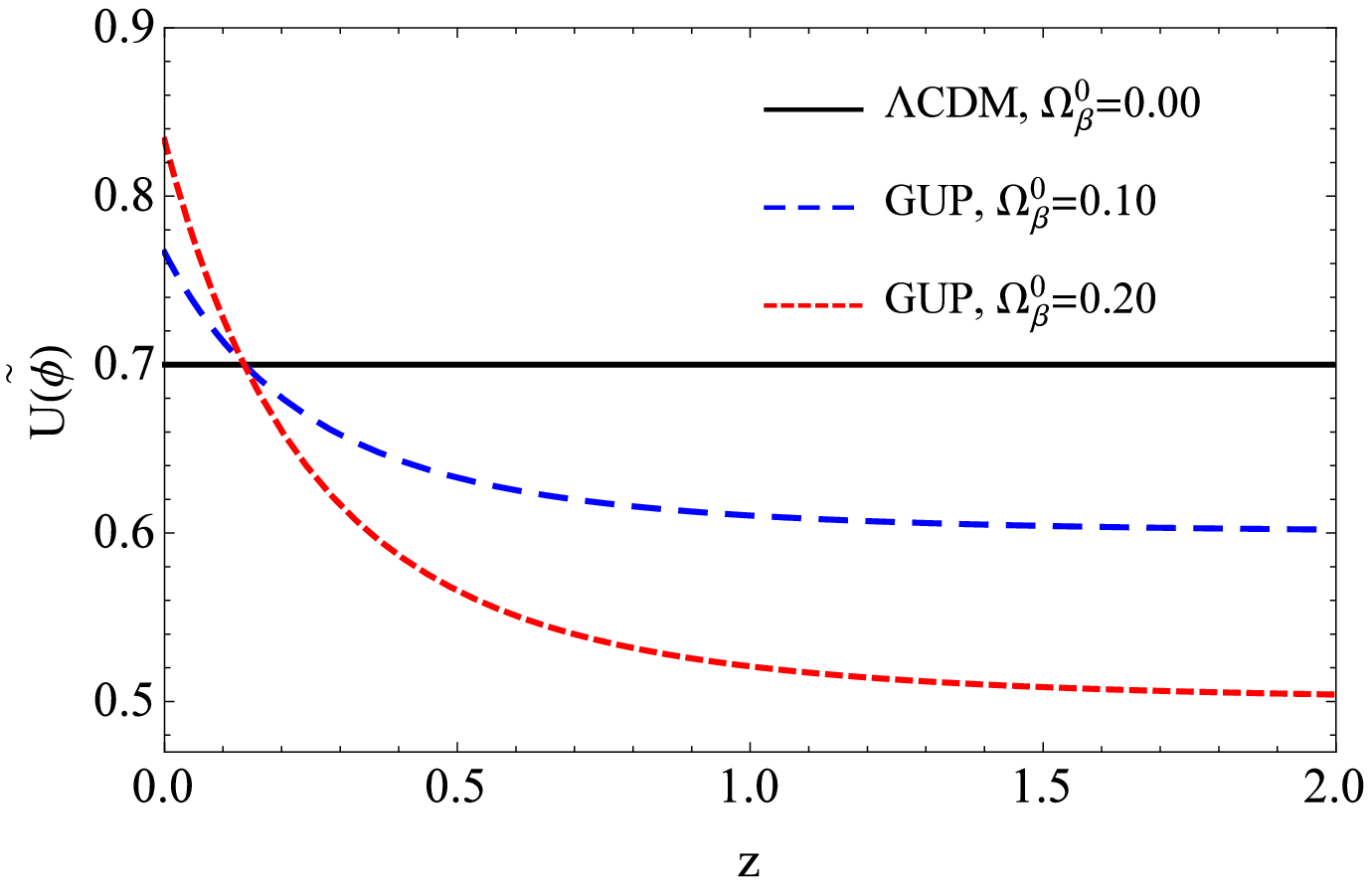}\label{figawesome_image2}}
\subfigure[]{\includegraphics[width=0.48\textwidth]{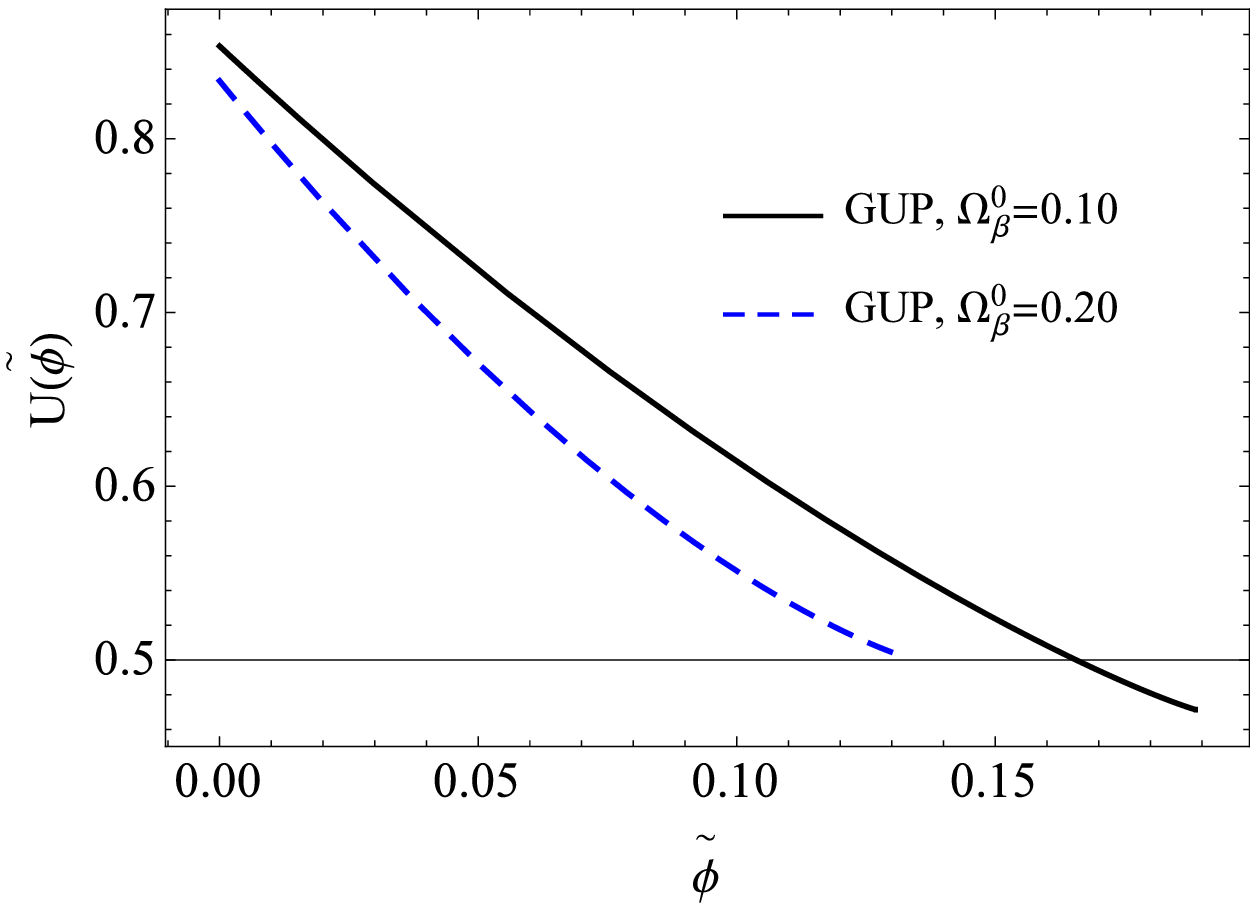}\label{figawesome_image3}}
\caption{\textit{panel a}: The behavior of the dimensionless potential $U(\tilde{\phi})$ respect to  the red-shift. \textit{panel b}: The behavior of the dimensionless potential $U(\tilde{\phi})$ respect to  the field $\tilde{\phi}$. We change the value of $\Omega^0_{\beta}$ while other parameters have been fixed to $\Omega^0_{DE}=0.70$ and $\Omega^0_{DM}=0.30$. }
\end{figure}

In the proposed dark energy model, energy density always is positive and the EoS is smaller { than $-1$, respect to the value of the parameter $\Omega^0_{\beta}$.} and,  $\dot{\phi}$ is a real filed. 
If, we want to apply such equation for our model, since $\omega_{{DE}}<-1$, it forces that $\dot{\psi}^{2}<0$, which means that the field must be imaginary. To remove this difficulty, we use the replacement $\dot{\psi}^{2}\longrightarrow -\dot{\Psi}^{2}$, therefore, the Tachyonic Lagrangian changes to
\begin{eqnarray}\label{new lag}
\mathcal{L}=-V(\Psi) \sqrt{1+ \dot{\Psi}^{2} }.
\end{eqnarray}
 On the other hand the potential term will be
\begin{eqnarray}\label{Phantom potential}
V(\phi)=\frac{1}{2} \left( 1- w_{\phi}\right)\rho_{\phi}.
\end{eqnarray}
To construct the scalar field, we assume that $\rho_{\phi}=\rho_{{DE}} $. According to the Eqs.~(\ref{dep}) and (\ref{omegam})
we obtain the kinetic term as
\begin{eqnarray}\label{phi tilde}
\phi'^{2}=\frac{3}{8 \pi G_{N}}  \frac{\epsilon(1+w_{\phi})\Omega_{\phi}}{(1+z)^{2} \left(\Omega_{{DM}}(z)+ \Omega_{\phi}(z)\right)},
\end{eqnarray}
which over-prime denotes differentiation  respect to red-shift. Substituting  ${\tilde{\phi}=\sqrt{\frac{8 \pi G_{N}}{3}} \phi}$, in  Eq.~(\ref{phi tilde}) changes to the form
\begin{eqnarray}
\tilde{\phi}'^{2}=\frac{\epsilon(1+w_{\phi})\Omega_{\phi}}{(1+z)^{2}\left(\Omega_{{DM}}(z)+ \Omega_{\phi}(z)\right)}.
\end{eqnarray}
Therefore,  we have
\begin{eqnarray}\label{tilde phi}
\frac{d\tilde{ \phi}}{dz}=\pm\sqrt{\frac{\epsilon(1+w_{\phi})\Omega_{\phi}}{(1+z)^{2} \left(\Omega_{{DM}}(z)+\Omega_{\phi}(z)\right)}}.
\end{eqnarray}
The sign of the above equation is arbitrary, from now on we choose the positive sign. By numerical analysis of Eq.~(\ref{tilde phi}), we show that the scalar field {dynamics} become important at the late times, as shown in Fig.~\ref{p-f}. On the other hand, using  Eq.~(\ref{Phantom potential}) we define a dimensionless scalar potential {according to}
\begin{eqnarray}\label{dimension-less potential}
U(\tilde{\phi})=\frac{8 \pi G_{N}}{3H^{2}_{0}}V(\tilde{\phi}),
\end{eqnarray}
which evolves with respect to the scalar field as shown in Fig.(\ref{figawesome_image2}).

From the above discussion, we conclude that the kinetic and potential terms at early times were constant which represent a cosmological constant, with the passage of time kinetic and potential terms start to evaluate with respect to time. So the Universe experience a smooth transition from dark matter dominated epoch to the late time dark energy dominated epoch.
 \subsection{Reconstruction of Tachyon Field}
Now, we use a similar approach in the previous section to construct another kind of scalar filed, namely Tachyon fields. Tachyon fields appear in string theory, using effective field theory \cite{Gibbons2002md}. The interaction between Tachyon field and dark matter were studied in Ref.~\refcite{Sheykhi2011fe}. If the field in this theory was an ordinary field, it is shown that such filed can be considered as a candidate for dark energy. It is shown that in the standard Tachyonic field, the equation of state changes between $-1 $ and $0$. Tachyon
field Lagrangian density is
\begin{eqnarray}
\mathcal{L}=-V(\psi)\sqrt{1-g^{\mu\nu}\partial_{\mu}\psi\partial_{\nu}\psi}.
\end{eqnarray}
It is important to notice that mass dimension of the scalar field is $M^{-1}$. If we consider that Tachyon field depends only on cosmic time, $t$, then the Lagrangian density becomes
\begin{eqnarray}
\mathcal{L}=-V(\psi) \sqrt{1- \dot{\psi}^{2} }.
\end{eqnarray}

\begin{figure}[t!]
\centering
\subfigure[]{\includegraphics[width=0.32\textwidth]{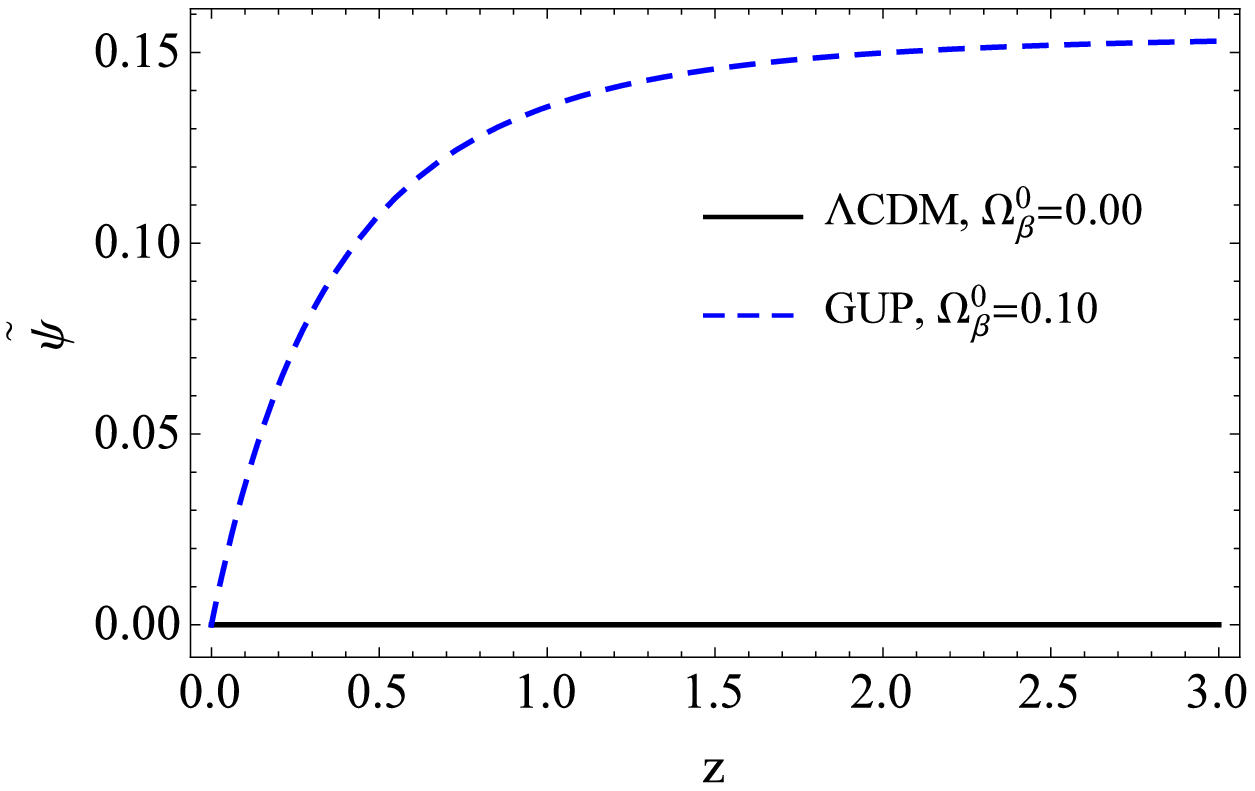}\label{Tachyon phi}}
\subfigure[]{\includegraphics[width=0.32\textwidth]{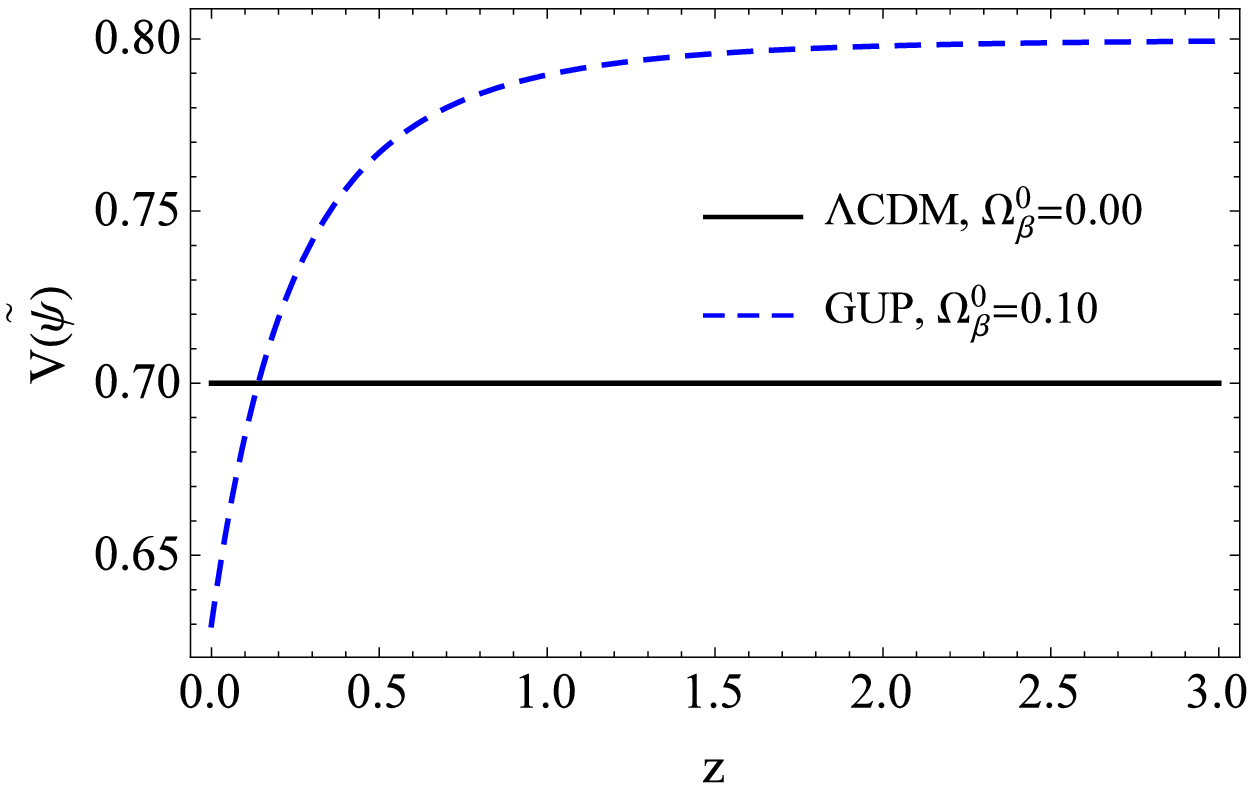}\label{u_potential_phi}}
\subfigure[]{\includegraphics[width=0.32\textwidth]{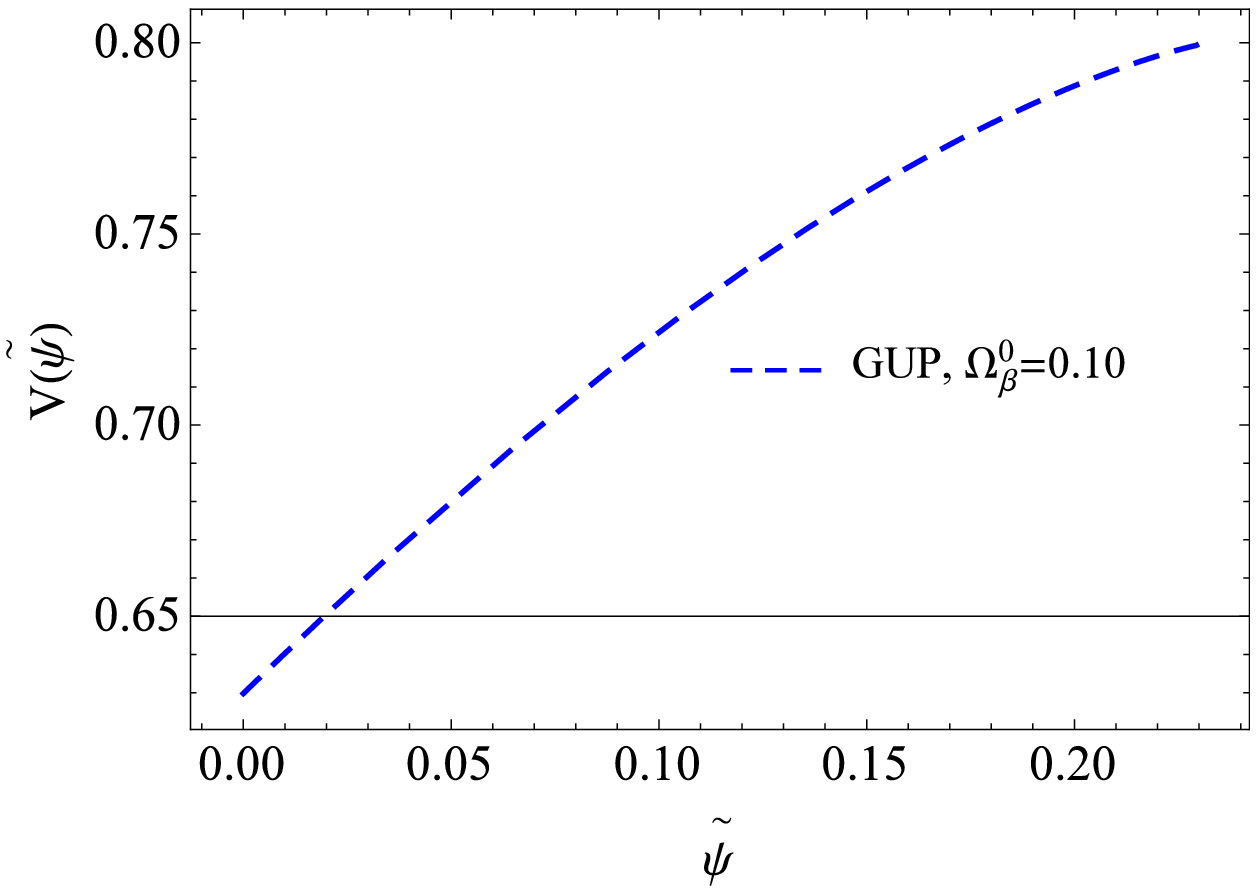}\label{u_potential_phi}}
\caption{\textit{panel a}:The evolution of  dimensionless Tachyon field, $\tilde{\psi}$, as a function of red-shift. We assume that $\tilde{\psi}(z=0)=0$. \textit{panel b}: The evolution of the dimensionless potential $V(\tilde{\psi})$
as a function of red-shift. \textit{panel c}: The behavior of dimensionless potential $V(\tilde{\psi})$
as a function of Tachyon filed $\tilde{\psi}$. }
\end{figure}
Hence, the energy density and the pressure associated to the Tachyon field will be
\begin{align}
\rho_{\psi}&=\frac{V(\psi)}{\sqrt{1-\dot{\psi}^{2}}}, \\
P_{\psi}&=-V(\psi) \sqrt{1- \dot{\psi}^{2} }.
\end{align}
Using above equations, the EoS parameter is
\begin{eqnarray}
w_{\psi}=\frac{P_{\psi}}{\rho_{\psi}}=-1+\dot{\psi}^{2}.
\end{eqnarray}

If, we want to apply such equation for our model, since $w_{{DE}}<-1$, it forces that $\dot{\psi}^{2}<0$, which means that the field must be imaginary. To remove this difficulty, we use the replacement $\dot{\psi}^{2}\longrightarrow -\dot{\Psi}^{2}$, therefore, the Tachyonic Lagrangian changes to
\begin{eqnarray}\label{new lag}
\mathcal{L}=-V(\Psi) \sqrt{1+ \dot{\Psi}^{2} }.
\end{eqnarray}
By considering Lagrangian density (\ref{new lag}) we can rewrite energy density and pressure as
\begin{align}
\rho_{\Psi}&=\frac{V(\Psi)}{\sqrt{1+\dot{\Psi}^{2}}},  \\
P_{\Psi}&=-V(\Psi) \sqrt{1+ \dot{\Psi}^{2} }.
\end{align}

So the EoS parameter for Tachyonic field becomes
\begin{eqnarray}
w_{\Psi}=\frac{P_{\Psi}}{\rho_{\Psi}}=-1- \dot{\Psi}^{2} ,
\end{eqnarray}
which satisfy the condition $w_{\Psi}<-1 $. Combining above equations, we obtain
\begin{eqnarray}
V(\Psi)=\pm \sqrt{-w_{\Psi}}\rho_{\Psi}.
\end{eqnarray}
Similar to the Phantom case, we choose the plus sign.

Now to establish a correspondence between our model and Tachyon field, we use the previous formula for EoS and energy density. If we define $\tilde{\psi}=H_{0} \psi $, the kinetic term rewrite as
\begin{eqnarray}
\frac{d \tilde{\psi}}{dz}= \frac{1}{1+z} \sqrt{\frac{-(1+w_{\psi})}{\Omega^{0}_{{DM}}(z)+\Omega_{\psi}(z)}}\, .
\end{eqnarray}
Again by numerical integration, we obtain the behavior of Tachyon field and potential as shown in Figs.~(\ref{Tachyon phi}) and (\ref{u_potential_phi}).
\section{Conclusion}
In this paper, we have applied the existence of minimal observable length to modify the classical field equations in standard cosmology. Considering a spatially flat Universe filled with the dark energy we obtain the effective equation of state for the dark side of the Universe containing dark energy and dark matter. We considered the interaction between dark energy and dark matter and showed that this interaction is a consequence of the effective EoS parameters which originated from the GUP parameter. Finally, we constructed two types of scalar fields for the said dynamical dark energy model, namely Phantom and Tachyon fields and compared their behavior during the evolution of the Universe.
Strictly speaking, we showed that the what is responsible for a dynamical dark energy model and interaction between dark components of the Universe could be the existence of minimal length in Universe.


\end{document}